\documentclass[english,final]{aipproc}
\usepackage[T1]{fontenc}
\usepackage[latin1]{inputenc}
\usepackage{graphicx}
\usepackage{amssymb}

\makeatletter

\layoutstyle{6x9}

\usepackage{babel}
\makeatother
\begin{document}

\newcommand{\missET}{{E_{T}\!\!\!\!\!\!\!/\:\:\,}}

\begin{flushright}Preprint ANL-HEP-CP-05-62, hep-ph/0507179\end{flushright}

\classification{}

\author{Anton V. Konychev}{address={Department of Physics, Indiana University, Bloomington, IN 47405-7105, U.S.A.}}

\author{Pavel M. Nadolsky}{address={High Energy Physics Division, Argonne National Laboratory, Argonne, IL 60439-4815, U.S.A.}}

\title{Universality of $q_{T}$ resummation\\
for electroweak boson production\footnote{Talk given by P. Nadolsky at the XIII International Workshop on Deep Inelastic Scattering (DIS~2005, April 27-May 1, 2005, Madison, WI, U.S.A.).}}

\begin{abstract}
We perform a global analysis of transverse momentum distributions
in Drell-Yan pair and $Z$ boson production in order to investigate
universality of nonperturbative contributions to the Collins-Soper-Sterman
resummed form factor. Our fit made in an improved nonperturbative
model suggests that the nonperturbative contributions follow universal
nearly-linear dependence on the logarithm of the heavy boson invariant
mass $Q$, which closely agrees with an estimate from the infrared
renormalon analysis. 
\end{abstract}
\maketitle
Transverse momentum distributions of heavy Drell-Yan lepton pairs,
$W$, or $Z$ bosons produced in hadron-hadron collisions present
an interesting example of factorization for multi-scale observables.
If the transverse momentum $q_{T}$ of the electroweak boson is much
smaller than its invariant mass $Q$, $d\sigma/dq_{T}$ at an $n$-th
order of perturbation theory includes large contributions of the type
$\alpha_{s}^{n}{\mathrm{ln}}^{m}(q_{T}^{2}/Q^{2})/q_{T}^{2}$ ($m=0,1\ldots2n-1$),
which must be summed through all orders of $\alpha_{s}$ to reliably
predict the cross section \cite{PP}. Such resummation is realized
in the Collins-Soper-Sterman (CSS) formalism \cite{CSS}, which describes
soft and collinear QCD radiation in a wide range of energies by introducing
a resummed form factor $\widetilde{W}(b)$ in impact parameter ($b$)
space. 

While the short-distance contributions ($b\lesssim1\mbox{ GeV}^{-1}$)
to the CSS form factor $\widetilde{W}(b)$ can be calculated in perturbative
QCD, long-distance nonperturbative contributions from $b>1\mbox{ GeV}^{-1}$
are not yet fully computable, even though their basic form can be
deduced from the infrared renormalon analysis \cite{KorS}. The factorization
theorem behind the CSS formalism predicts that the nonperturbative
contributions are universal in unpolarized Drell-Yan-like and semi-inclusive
DIS processes. Consequently the function ${\mathcal{F}}_{NP}(b,Q)$
that describes the nonperturbative terms can be constrained in a global
fit to the hadronic $q_{T}$ data, just as the $k_{T}$-integrated
parton densities are constrained with the help of inclusive scattering
data. ${\mathcal{F}}_{NP}(b,Q)$ must be known precisely in order
to successfully measure the $W$ boson mass, because uncertainties
in ${\mathcal{F}}_{NP}(b,Q)$ may affect the measured value of $M_{W}$
at the level comparable to the targeted accuracy of the measurement,
$\delta M_{W}\approx30$ MeV at the Tevatron and 15 MeV at the LHC.
It is therefore interesting to investigate if ${\mathcal{F}}_{NP}(b,Q)$
found in the $q_{T}$ fit is consistent with the universality hypothesis,
and whether its preferred form is compatible with the renormalon analysis. 

These issues were explored recently in Ref.~\cite{KN2005}, where
a global analysis of $q_{T}$ data from fixed-target Drell-Yan pair
production and Tevatron $Z$ boson production was performed in the
context of an improved model for the nonperturbative contributions.
Although ${\mathcal{F}}_{NP}(b,Q)$ primarily parametrizes the {}``power-suppressed''
terms, \textit{i.e.}, terms proportional to positive powers of $b$,
its form found in the fit is correlated with the assumed behavior
of the leading-power terms (logarithmic in $b$ terms) at $b<2\mbox{ GeV}^{-1}$.
The exact behavior of $\widetilde{W}(b)$ at $b>2\mbox{ GeV}^{-1}$
is of reduced importance, as $\widetilde{W}(b)$ is strongly suppressed
at such $b$. For these reasons, we closely followed the procedure
of the previous global $q_{T}$ analysis \cite{BLNY}, while paying
close attention to the model of the leading-power terms at perturbative
and moderately nonperturbative transverse distances, $b<2\mbox{ GeV}^{-1}$.

The large-$b$ contributions were introduced by using the $b_{*}$
model \cite{CSS}, as\begin{equation}
\widetilde{W}(b)=\widetilde{W}_{pert}(b_{*})\, e^{-{\mathcal{\mathcal{F}}}_{NP}(b,Q)}.\label{W2}\end{equation}
Here $\widetilde{W}_{pert}(b_{*})$ is the perturbative part of $\widetilde{W}(b)$,
\textit{i.e.}, its leading-power part evaluated at a finite order
of $\alpha_{s}$. $\widetilde{W}_{pert}(b_{*})$ depends on the variable
$b_{*}\equiv b/(1+b^{2}/b_{max}^{2})^{1/2}$ and serves as an approximation
for all leading-power terms. Its shape is varied at all $b$ by adjusting
a single parameter $b_{max}$. The $b_{*}$ model with a relatively
low $b_{max}=0.5\mbox{ GeV}^{-1}$ was a choice of the previous $q_{T}$
fits \cite{BLNY,DWSLYERV}. However, it is natural to consider $b_{max}$
above 1$\mbox{ GeV}^{-1}$ in order to avoid \emph{ad hoc} modifications
of $\widetilde{W}_{pert}(b)$ in the $b$ region where perturbation
theory is still applicable. In Ref.~\cite{KN2005}, we proposed a
modification in the $b_{*}$ model that allowed us to increase $b_{max}$
at least up to $\approx3\mbox{ GeV}^{-1}$, while preserving correct
resummation of the large logarithms at small $b$ and numerical stability
of the Fourier-Bessel transform. If a very large $b_{max}$ comparable
to $1/\Lambda_{QCD}$ is taken, $\widetilde{W}_{LP}(b)$ essentially
coincides with $\widetilde{W}_{pert}(b)$, extrapolated to large $b$
by using the known, although not always reliable, dependence of $\widetilde{W}_{pert}(b)$
on $\ln{b}$. Hence, the new prescription can be also used to test
viability of extrapolation of $\widetilde{W}_{pert}(b)$ to large
$b$, reminiscent of similar extrapolations introduced in the alternative
models~\cite{QZ,KSV}. 

\begin{figure}
\includegraphics[%
  clip,
  width=0.49\columnwidth]{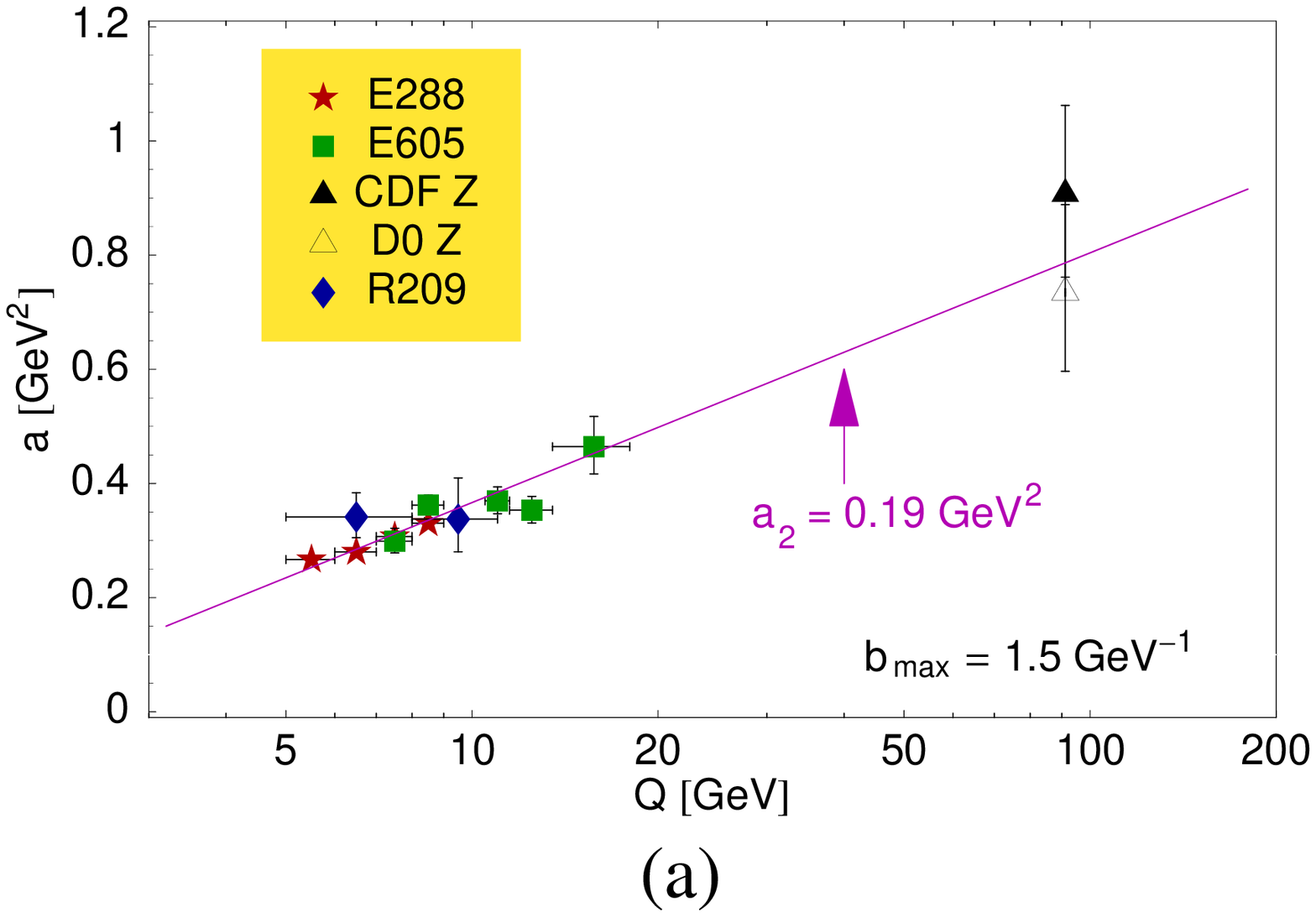}\includegraphics[%
  clip,
  width=0.49\columnwidth]{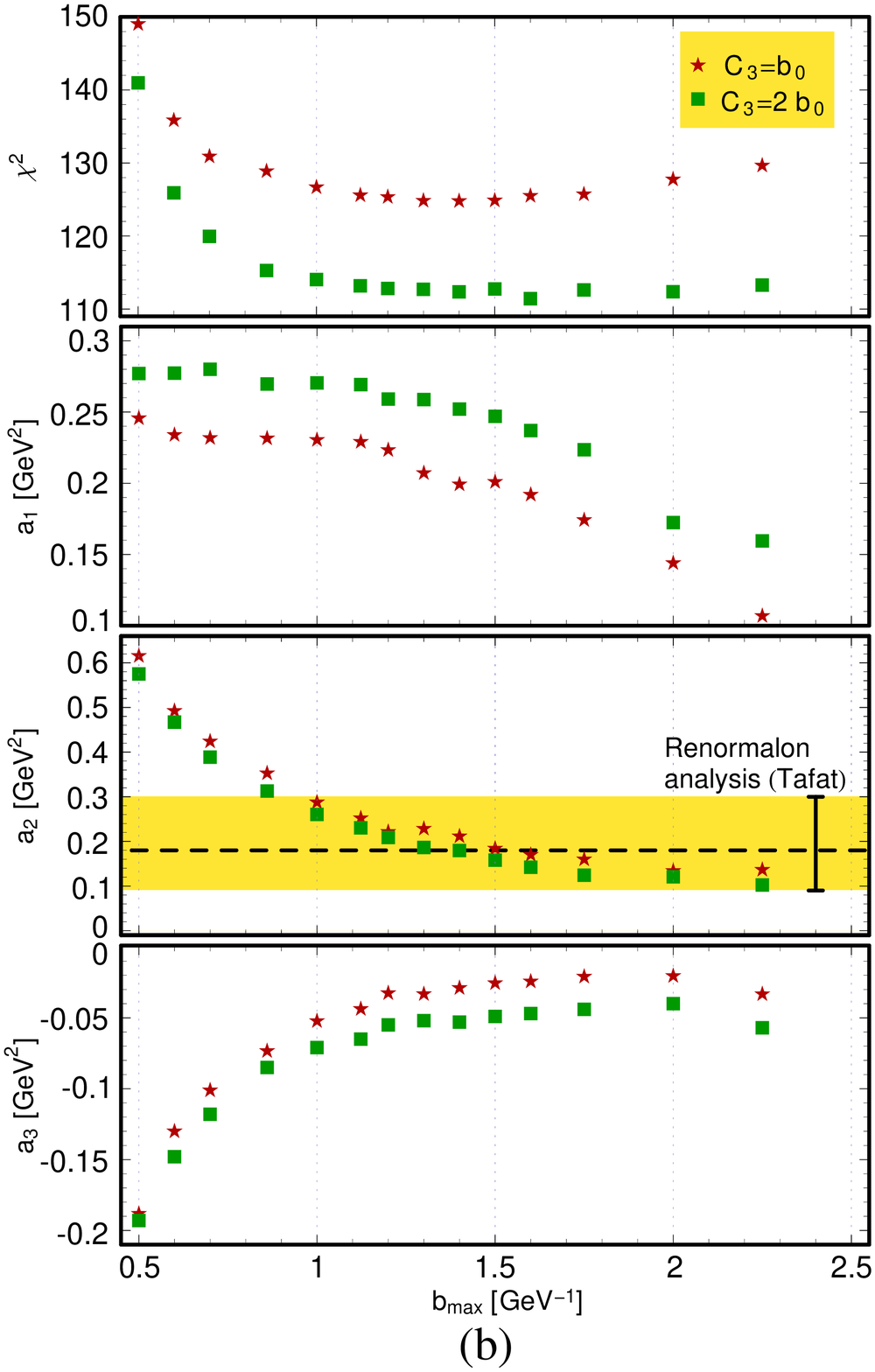}

\caption{(a) The best-fit values of $a(Q)$ obtained in independent scans
of $\chi^{2}$ for the contributing experiments. The vertical error
bars correspond to the increase of $\chi^{2}$ by unity above its
minimum in each $Q$ bin. The slope of the line is equal to the central-value
prediction from the renormalon analysis \cite{Tafat}. (b) The best-fit
$\chi^{2}$ and coefficients $a_{1}$, $a_{2}$, and $a_{3}$ in ${\mathcal{F}}_{NP}(b,Q)$
for different values of \protect $b_{max}$. The size of the symbols
approximately corresponds to $1\sigma$ errors for the shown parameters.\label{fig:aQvsQ}\vspace{-0.5cm}}
\end{figure}

Following the renormalon analysis and Ref.~\cite{BLNY}, we assumed
a Gaussian form of the nonperturbative function, ${\cal F}_{NP}(b,Q)\equiv a(Q)b^{2},$
with\begin{equation}
a(Q)\equiv a_{1}+a_{2}\mathrm{ln}\left[Q/(3.2\mbox{ GeV})\right]+a_{3}\mathrm{ln}\left[100x_{1}x_{2}\right].\label{KNform}\end{equation}
The dependence of ${\mathcal{F}}_{NP}$ on $\ln Q$ is a consequence
of renormalization-group invariance of the soft-gluon radiation. The
coefficient $a_{2}$ of the $\ln Q$ term has been related to the
vacuum average of the Wilson loop operator and evaluated within lattice
QCD as $0.19_{-0.09}^{+0.12}\mbox{ GeV}^{2}$\nolinebreak[4] \cite{Tafat}.
To see if the universal Gaussian behavior is consistent with the data,
we first examined the values of $a(Q)$ that are independently preferred
by each bin of $Q$ in 5 examined experimental data sets. Fig.~\ref{fig:aQvsQ}(a)
shows the best-fit values of $a(Q)$ obtained in independent fits
to the data in each bin of $Q$ for $b_{max}=1.5\mbox{ GeV}^{-1}$.
The best-fit $a(Q)$ follow a nearly linear dependence on $\ln Q$,
and the slope $a_{2}\equiv da(Q)/d(\ln Q)$ is close to the renormalon
analysis expectation of $~0.19\mbox{ GeV}^{2}$ \cite{Tafat}. Such
nearly linear behavior of $a(Q)$ is observed in the entire range
$b_{max}=1-2\mbox{ GeV}^{-1}$, and it less pronounced at $b_{max}$
outside of the interval 1-2$\mbox{ GeV}^{-1}$. Since the best-fit
$a(Q)$ in each $Q$ bin are essentially independent, we conclude
that the data support the universality of ${\mathcal{F}}_{NP}$, when
$b_{max}$ lies in the range $1-2\mbox{ GeV}^{-1}$. In addition,
each experimental data set individually prefers a nearly quadratic
dependence on $b$, ${\mathcal{F}}_{NP}=a(Q)b^{2-\beta}$, with $|\beta|<0.5$
in all experiments.

Next, we performed a simultaneous fit of our model to all the data.
Fig.~\ref{fig:aQvsQ}(b) shows the dependence of the best-fit $\chi^{2},$
$a_{1},$ $a_{2}$, and $a_{3}$ on $b_{max}$. As $b_{max}$ is increased
above $0.5\mbox{ GeV}^{-1}$ assumed in the studies \cite{BLNY,DWSLYERV},
$\chi^{2}$ rapidly decreases, becomes relatively flat at $b_{max}=1-2\mbox{ GeV}^{-1}$,
and grows again at $b_{max}>2\mbox{ GeV}^{-1}$. The global minimum
of $\chi^{2}$ is reached at $b_{max}\approx1.5$ GeV$^{-1}$, where
all data sets are described equally well, without major tensions among
the five experiments. The magnitudes of $a_{1}$, $a_{2}$, and $a_{3}$
are reduced when $b_{max}$ increases from $0.5$ to $1.5\mbox{ GeV}^{-1}$.
In the whole range $1\leq b_{max}\leq2$ GeV$^{-1}$, $a_{2}$ agrees
with the renormalon analysis estimate. The coefficient $a_{3}$, which
parametrizes deviations from the linear $\ln Q$ dependence, is considerably
smaller ($<0.05$) than both $a_{1}$ and $a_{2}$ ($\sim0.2$). This
behavior supports the conjecture in \cite{QZ} that $a_{3}$ is small
if the exact form of $\widetilde{W}_{pert}(b)$ is maximally preserved.

The preference for the values of $b_{max}$ between $1$ and $2\mbox{ \, GeV}^{-1}$
indicates, first, that the data do favor the extension of the $b$
range where all leading-power terms are approximated by their finite-order
expression $\widetilde{W}_{pert}(b)$. In $Z$ boson production, this
region extends up to $3-4\mbox{ GeV}^{-1}$ as a consequence of the
strong suppression of the large-$b$ tail by the Sudakov exponent.
The fit to the $Z$ data is actually independent of $b_{max}$ within
the experimental uncertainties for $b_{max}>1\mbox{ GeV}^{-1}$. In
the low-$Q$ Drell-Yan process, the continuation of $b\widetilde{W}_{pert}(b)$
far beyond $b\approx1\mbox{ GeV}^{-1}$ is disfavored because of large
higher-order corrections to $b\widetilde{W}_{pert}(b)$ at $b$ around
$1.5\mbox{ GeV}^{-1}$. To summarize, the extrapolation of $\widetilde{W}_{pert}(b)$
to $b>\nolinebreak1.5\mbox{ GeV}^{-1}$ is disfavored by the low-$Q$
data sets, if a purely Gaussian form of ${\mathcal{F}}_{NP}$ is assumed.
The Gaussian approximation is adequate, on the other hand, in the
$b_{*}$ model with $b_{max}$ in the range $1-2\mbox{ GeV}^{-1}$.

In $Z$ boson production, our best-fit $a(M_{Z})=0.85\pm0.10\mbox{ GeV}^{2}$
agrees with $0.8\mbox{ GeV}^{2}$ found in the extrapolation-based
models \cite{QZ,KSV}, and it is about a third of $2.7\mbox{ GeV}^{2}$
predicted by the BLNY parametrization. In the low-$Q$ Drell-Yan case,
our $a(Q)=0.2-0.4\mbox{ GeV}^{2}$ is close to the average $\langle a\rangle=0.19-0.28\mbox{ GeV}^{2}$
in four $Q$ bins of the E288 and E605 data found in the model \cite{QZ}.
To describe the low-$Q$ data, Ref.~\cite{QZ} allowed a large discontinuity
in the first derivative of $\widetilde{W}(b)$ at $b$ equal to the
separation parameter $b_{max}^{QZ}=0.3-0.5\mbox{ GeV}^{-1}$, where
switching from the exact $\widetilde{W}_{pert}(b)$ to its extrapolated
form occurs. In the revised $b_{*}$ model, such discontinuity does
not happen, and $\widetilde{W}_{LP}(b)$ is closer to the exact $\widetilde{W}_{pert}(b)$
in a wider $b$ range than in Ref.~\cite{QZ}.

The best-fit parameters in ${\mathcal{F}}_{NP}$ found in the new
model are quoted in Ref.~\cite{KN2005}. The global fit places stricter
constraints on ${\mathcal{F}}_{NP}$ at $Q=M_{Z}$ than the Tevatron
Run-1 $Z$ data alone. Theoretical uncertainties from a variety of
sources may be substantial in the low-$Q$ Drell-Yan process, which
is indicated, in particular, by the dependence of the agreement with
the low-$Q$ data on an arbitrary factorization scale $C_{3}$ in
$\widetilde{W}_{pert}(b)$. The low-$Q$ uncertainties do not substantially
affect predictions at the electroweak scale. The ${\mathcal{O}}(\alpha_{s}^{2})$
corrections and scale dependence are smaller in $W$ and $Z$ production,
and, in addition, the term $a_{2}\ln Q$, which arises from the soft
factor ${\mathcal{S}}(b,Q)$ and dominates ${\mathcal{F}}_{NP}$ at
$Q=M_{Z}$, shows little variation with $C_{3}$. Consequently, the
revised $b_{*}$ model with $b_{max}\approx1.5\mbox{ GeV}^{-1}$ increases
our confidence in the transverse momentum resummation at electroweak
scales by exposing the soft-gluon origin and universality of the dominant
nonperturbative contributions at collider energies.

\begin{theacknowledgments}We express our gratitude to C.-P. Yuan
for his crucial help with the setup of the fitting program. This work
was supported in part by the US Department of Energy, High Energy
Physics Division, under Contract W-31-109-ENG-38, and by the U.S.
National Science Foundation under grants PHY-0100348 and PHY-0457219.\end{theacknowledgments}
\end{document}